# Ghost Imaging with Atoms

R. I. Khakimov, B. M. Henson, D. K. Shin, S. S. Hodgman, R. G. Dall, K. G. H. Baldwin and A. G. Truscott*

*Research School of Physics and Engineering, Australian National University, Canberra 0200, Australia*

(Dated: 01 July 2016)

Ghost imaging is a technique – first realized in quantum optics [1, 2] – in which the image emerges from cross-correlation between particles in two separate beams. One beam passes through the object to a bucket (single-pixel) detector, while the second beam's spatial profile is measured by a high resolution (multi-pixel) detector but never interacts with the object. Neither detector can reconstruct the image independently. However, until now ghost imaging has only been demonstrated with photons. Here we report the first realisation of ghost imaging of an object using massive particles. In our experiment, the two beams are formed by correlated pairs of ultracold metastable helium atoms [3], originating from two colliding Bose-Einstein condensates (BECs) via *s*-wave scattering [4, 5]. We use the higher-order Kapitza-Dirac effect [6] to generate the large number of correlated atom pairs required, enabling the creation of a ghost image with good visibility and sub-millimetre resolution. Future extensions could include ghost interference as well as tests of EPR entanglement [7] and Bell's inequalities [8].

Ghost imaging with light is a surprising, counter-intuitive phenomenon, which allows the image of an object to be reconstructed from the spatio-temporal properties of a beam that never interacts with it. There is also no requirement for any spatial measurement of the light beam that does pass through the object, merely the temporal detection of the associated total light intensity. Temporal correlations between the two beams can then be used to reconstruct the image.

Since the first experimental realisation of ghost imaging with light more than twenty years ago [9, 10], following the original proposals some years previously [11, 12], there has been an explosion of interest in the technique [1, 2]. Ghost imaging has since found applications in areas ranging from environmental sensing [13] to cryptography [14]. Extension of these techniques has enabled ghost imaging in the X-ray domain [15], 3-D ghost imaging [16] and even temporal ghost imaging with applications to improved telecommunications [17]. Ghost imaging has produced sub-shot-noise images of weakly absorbing objects [18] and for low-light-level imaging, when the number of registered photons per image pixel is less than one, ghost imaging has been shown to outperform conventional imaging in terms of contrast [19].

In the early stages there was considerable debate as to whether ghost imaging was a semi-classical or a quantum-optical phenomenon. This debate was resolved [1, 2], and showed that ghost imaging can be realised either using a source of thermal [20, 21]/pseudothermal [22] photons, or correlated photon pairs (bi-photons) such as created in spontaneous parametric down conversion (SPDC) [21]. Thermal/pseudothermal ghost imaging is not a quantum effect and can be explained semi-classically, whereas SPDC ghost imaging requires quantum theory to describe its quantitative performance and can demonstrate violation of Bell's inequalities [8].

To date this rich history of experimental ghost imaging has been achieved exclusively with photons. However, until now there has been no realisation of ghost imaging using massive particles. As well as demonstrating complementarity for this phenomena using matter waves, realising ghost imaging with atoms is a potential precursor to experiments that test fundamental concepts in quantum mechanics with massive particles, such as ghost interferometry, Einstein-Podolsky-Rosen entanglement and Bell's inequalities [7, 8]. Further, there are potential applications such as real-time, *in situ* control of atom lithography while imaging the deposition remotely *via* the spatial (multi-pixel) detector as the process is taking place.

There are two main challenges to ghost imaging with massive particles: first, the need for a source of particles with the required correlation properties; second, the need for a sufficiently high flux to provide adequate measurement statistics. The advent of atom cooling techniques that enable the creation and manipulation of BECs allows such a source to be created. A schematic of ghost imaging using correlated colliding atom pairs is illustrated in Fig. 1.

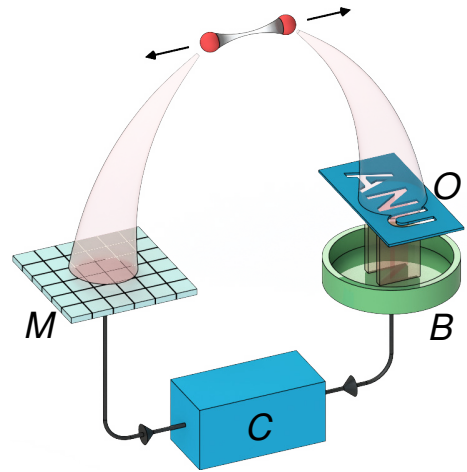

FIG. 1. **Schematic for atomic ghost imaging.** Correlated pairs of atoms created in a collision form two beams. One beam passes through the object to be imaged (*O*) and their arrival times are detected by a bucket detector (*B*). The second beam never interacts with the object, but is detected with full temporal and spatial resolution by a multi-pixel detector (*M*). A correlator (*C*) then reconstructs the image of the object.

As with optics, two types of atom sources can be envisaged. Like thermal light sources, a thermal ensemble of ultracold atoms possesses second- ($1 < g^{(2)} < 2$) and higher-order correlations, as demonstrated *via* the Hanbury Brown and Twiss effect [23–25]. However, the flux of correlated atom pairs is usually very low. Alternatively, the degree of correlation can be enhanced by exploiting the atom equivalent of bi-photon sources, whereby *s*-wave collisions between ultracold atoms [4, 5] can in principle yield a high degree of correlation ($g^{(2)} \gg 2$) between two atoms. Even then, the requirements on the flux to realise a ghost imaging experiment are quite severe.

To overcome this we have developed a new technique to enhance the number of correlated atom pairs, while maintaining a high degree of correlation and therefore signal-to-noise ratio. We use higher-order Kapitza-Dirac scattering [6, 26, 27] to produce multiple distinct sources of $s$-wave scattered atoms in each experimental run. This allows more than a 10-fold increase in the data acquisition rate which makes the experiment feasible – the data we present here is the equivalent of around three weeks of full-time data acquisition.

The experiments start with a Bose-Einstein condensate (BEC) of helium atoms in the metastable ($2^3S_1$) state. This state enables single atom detection with high efficiency because of the large internal energy of the atoms [3].

We magnetically trap a BEC of $\sim 10^6$ $^4$He* atoms in the $m_J = +1$ sublevel with no discernible thermal fraction (see [28, 29] for details). To produce an $s$-wave halo, we collide atoms in the BEC in two steps: (i) first we outcouple nearly all atoms from the trap via a two-photon Raman transition to the magnetically insensitive sublevel $m_J = 0$, and then (ii) Bragg scatter the cloud into multiple momentum modes. This last step introduces a relative momentum difference between atoms in the different Bragg orders, which then generate a series of $s$-wave scattering halos via binary collisions between atoms in different orders [29]. We employ the same laser beams for both Raman and Bragg pulses, with the latter Fourier-broadened to ensure we are in the Kapitza-Dirac regime, which populates multiple diffraction orders. Both Raman and Bragg pulses propagate along the $(\mathbf{e}_x \pm \mathbf{e}_z)/\sqrt{2}$ directions as shown in Fig. 2a(I). This results in momentum transfer along the vertical axis $\mathbf{e_z}$ with the momentum difference between any two adjacent diffracted orders $\hbar\Delta\mathbf{k} = \pm\sqrt{2}\hbar k_0 \mathbf{e_z}$, where $\hbar k_0$ is a single photon recoil, $k_0 = 2\pi/\lambda$, and $\lambda = 1083.1979$ nm is the photon wavelength. In $k$-space associated with the center-of-mass reference frame, each $s$-wave scattering halo comprises atoms on a sphere of radius $k_r = \Delta k/2 = k_0/\sqrt{2}$, which reflects energy and momentum conservation, as shown in Fig. 2A(II).

After the collision, the expanding halo falls $\sim 850$ mm under gravity onto a multi-channel plate and delay-line detector, which records the arrival time and positions $(t,x,y)$ of individual atoms. We then transform this 3D information into the momenta $(k_x, k_y, k_z)$ of individual atoms, which is plotted in Fig. 2b to show the eleven halos produced in the collision process. The dark clouds represent BEC's in different Bragg orders $l$ with an $s$-wave halo situated between each of the corresponding orders $(l, l+1)$. The center of each halo is on the momentum transfer axis $\mathbf{e_z}$. Also clearly visible are the portions of larger-diameter halos from non-adjacent orders $(l, l+2)$, etc., as well as halos formed by single-photon spontaneous scattering from the Raman/Bragg laser. The halo populations follow that of Bragg orders in the Kapitza-Dirac effect, and as a consequence, the average number of atoms per halo decreases from $\sim 250$ for $(l, l+1) = (0, +1)$ to $\sim 50$ for $(+5, +6)$.

Ghost imaging is demonstrated by placing a thin metal mask 10 mm above the detector, which covers a portion of the detector's surface such that only a fraction of the $s$-wave halo (containing at most one atom from each correlated pair) passes through the mask. The rest of the atoms are detected directly, without any interaction with the object. The detector is artificially split into two regions: the bucket region, masked with the object we want to image, and the multi-pixel region, where the atoms are detected with full spatial and temporal resolution. On the bucket portion of the detector, only the time of arrival $t$ is used for ghost image reconstruction, whereas for the multi-pixel detector we retain a full $(t, x, y)$ set. The ghost image is then reconstructed in momentum space using coincidence-counting between atoms in the multi-pixel port and the bucket port. Combining the independent images from halos in the different Bragg orders results in the ghost image at the bottom of Fig. 2c. The ghost image clearly resolves the object, which is a 15.5 mm wide mask of the letters "ANU" with line widths $\sim 0.75$ mm.

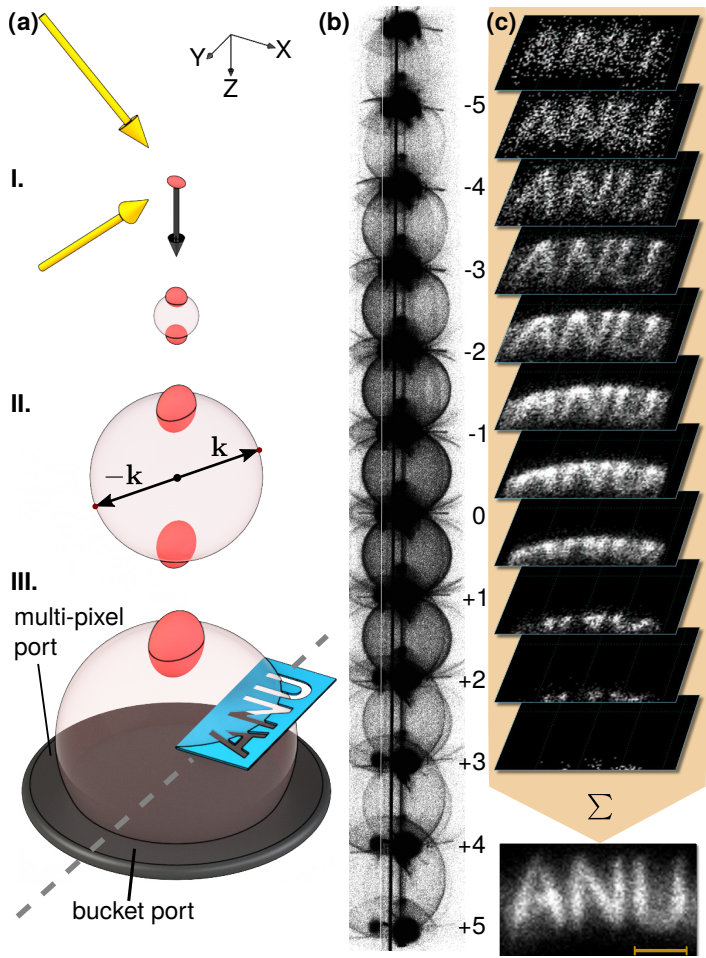

FIG. 2. **Schematic of the experiment and resulting ghost image.** (a) The experiment starts (I) by using Bragg laser beams (yellow arrows) to split the trapped BEC cloud into different momentum states (for simplicity, only the 1st Bragg order is shown as the clouds evolve in time). Binary atomic collisions populate an $s$-wave scattering halo with correlated pairs of opposite momenta, which then expand as they fall under gravity. (II) In the momentum space associated with the center-of-mass reference frame the BECs are situated at opposite poles of the $s$-wave sphere. (III) Some of the halo atoms pass through a mask placed 10 mm above the "bucket" port of the single atom detector, while their diametrically opposing counterparts are registered by the "multi-pixel" port. (b) Experimental data from 2,000 individual experimental runs showing the eleven halos produced in the collision process, where the individual atom counts are reconstructed in 3D momentum space. Bragg scattering in the Kapitza-Dirac effect produces 12 diffraction orders $-6 \cdots +5$ along $\mathbf{e}_z$. Collisions between each pair of adjacent orders result in 11 independent scattering halos. (c) Individual ghost images from each of the halos from 68,835 experimental runs are combined to form the final image (bottom). The scale bar in the final image is 5 mm. Because of the difference in absolute velocities for different Bragg orders, the halos which land first only cover a fraction of the image.

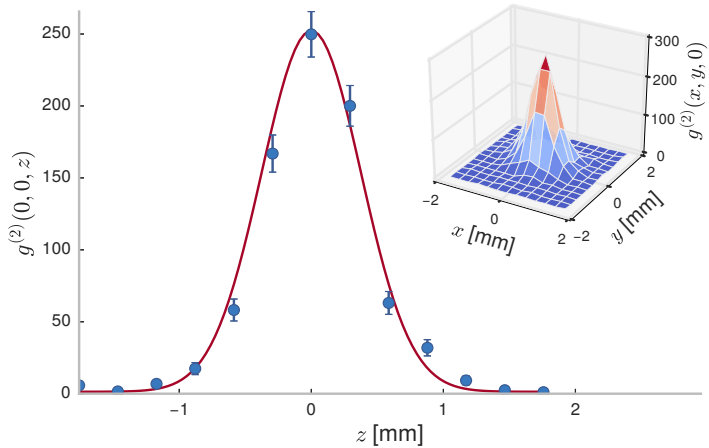

FIG. 3. **Cross-correlation function.** The main plot shows $g^{(2)}(0,0,z)$ as a function of vertical coordinate only. The solid line is a Gaussian fit, which has an rms width $\sigma_z = 0.37$ mm, corresponding to the correlation length. Error bars represent the standard error of the mean. The inset shows $g^{(2)}(x,y,0)$ for the experimental data.

Following [4], we characterise the correlations between atoms with almost equal but opposite momenta $\mathbf{k}_1 = -\mathbf{k}_2 + \Delta \mathbf{k}$ by constructing a two-particle cross-correlation function $g^{(2)}(\Delta k_x, \Delta k_y, \Delta k_z)$ [29]. Fig. 3 shows $g^{(2)}$ for the halo from Bragg orders $l = +3, +4$. The key factor limiting the image resolution is the finite width of $g^{(2)}$ – the correlation length.

A smaller average number of counts per halo is beneficial, as it leads to a higher peak $g^{(2)}$ [4] and, consequently, lower probability of registering false coincidence counts, which contribute to the ghost image background. This is why multi-order Kapitza-Dirac scattering is crucial for this type of experiment, as it allows a large number of relatively dilute s-wave halos to be populated and detected in a single experimental run, thereby significantly decreasing the required experimental run-time.

In the $xy$-plane of the detector, the absolute image resolution is limited by the real-space widths of $g^{(2)}(\Delta x, \Delta y, 0)$, while the spread along $z$ contributes to background counts during image reconstruction. The $g^{(2)}(\Delta x, \Delta y, \Delta z)$ data in Fig. 3 is fitted with a 3D Gaussian function $f(\Delta x, \Delta y, \Delta z) = 1 + A \exp(-\Delta x^2/2\sigma_x^2 - \Delta y^2/2\sigma_y^2 - \Delta z^2/2\sigma_z^2)$. This yields rms widths (correlation lengths) of $\{\sigma_{x,y,z}\} = \{0.43, 0.39, 0.37\}(\pm 0.01)$ mm, corresponding to the momentum spread of the halo.

The resolution of our ghost imaging setup can also be estimated by analysing the image of a known shape, which we do in Fig. 4a for the bars of the "U", which are 0.77 mm wide. Fitting the convolution of the object with a one-dimensional Gaussian point spread function (PSF) gives the rms width of the PSF to be $0.40(\pm 0.03)$ mm along the $x$-direction. This is in good agreement with the value for $\sigma_x = 0.43(\pm 0.01)$ mm obtained from the analysis of the $g^{(2)}$ widths.

In Fig. 4b we plot the visibility $V = (I - B)/(I + B)$ of the ghost image [19], where $I$ is the average ghost image intensity and $B$ is the average background. We show a build-up in the ghost image visibility as we cumulatively add contributions from different Bragg orders [29]. The optimal final image utilises 9 halos and has a visibility $V \approx 35\%$.

In conclusion, by using higher-order Bragg scattering in the Kapitza-Dirac effect, we are able to extend the technique of pair production via s-wave scattering and demonstrate more than a 10-fold increase in the number of correlated pairs available for each single experimental run. Further, because we have created a source of strongly correlated atom pairs, by analogy with SPDC we are able to generate very high second order $g^{(2)}$ correlation values greatly exceeding the maximum thermal value of 2, with maxima in our case reaching $g^{(2)}(0,0,0) \approx 250$. Using this source of correlated twin beams we perform ghost imaging with atoms, achieving good visibility and showing that the sub-millimetre resolution of the image is limited by the two-particle correlation function of the atomic momenta. This demonstration opens up a number of exciting possibilities for fundamental tests of quantum mechanics using massive particles, such as tests of EPR entanglement [7] and Bell's inequalities [8]. In particular, the properties of our source would overcome the major challenges (such as pair identification) outlined in the proposal from the Zeilinger group [7].

---

* To whom correspondence should be addressed; E-mail: andrew.truscott@anu.edu.au
[1] B. I. Erkmen and J. H. Shapiro, Adv. Opt. Photon. **2**, 405 (2010).
[2] J. H. Shapiro and R. W. Boyd, Quantum Information Processing **11**, 949 (2012).
[3] W. Vassen, C. Cohen-Tannoudji, M. Leduc, D. Boiron, C. I. Westbrook, A. Truscott, K. Baldwin, G. Birkl, P. Cancio, and M. Trippenbach, Reviews of Modern Physics **84**, 175 (2012), arXiv:1110.1361.
[4] A. Perrin, H. Chang, V. Krachmalnicoff, M. Schellekens, D. Boiron, A. Aspect, and C. I. Westbrook, Phys. Rev. Lett. **99**, 150405 (2007).
[5] J. C. Jaskula, M. Bonneau, G. B. Partridge, V. Krachmalnicoff, P. Deuar, K. V. Kheruntsyan, a. Aspect, D. Boiron, and C. I. Westbrook, Physical Review Letters **105**, 1 (2010), arXiv:1008.0845.
[6] P. L. Gould, G. A. Ruff, and D. E. Pritchard, Phys. Rev. Lett. **56**, 827 (1986).
[7] J. Kofler, M. Singh, M. Ebner, M. Keller, M. Kotyrba, and A. Zeilinger, Physical Review A - Atomic, Molecular, and Optical Physics **86**, 1 (2012), arXiv:1206.2141.
[8] B. Jack, J. Leach, J. Romero, S. Franke-Arnold, M. Ritsch-Marte, S. M. Barnett, and M. J. Padgett, Physical Review Letters **103**, 1 (2009).
[9] T. B. Pittman, Y. H. Shih, D. V. Strekalov, and A. V. Sergienko, Phys. Rev. A **52**, R3429 (1995).
[10] D. V. Strekalov, A. V. Sergienko, D. N. Klyshko, and Y. H. Shih, Phys. Rev. Lett. **74**, 3600 (1995).
[11] D. Klyshko, JETP **67**, 1131 (1988).
[12] A. V. Belinskii and D. N. Klyshko, JETP **78**, 259 (1994).
[13] N. D. Hardy and J. H. Shapiro, Phys. Rev. A **87**, 023820 (2013).
[14] S. Yuan, J. Yao, X. Liu, X. Zhou, and Z. Li, Optics Communications **365**, 180 (2016).
[15] H. Yu, R. Lu, S. Han, H. Xie, G. Du, T. Xiao, and D. Zhu, ArXiv e-prints (2016), arXiv:1603.04388 [physics.optics].
[16] B. Sun, M. P. Edgar, R. Bowman, L. E. Vittert, S. Welsh, a. Bowman, and M. J. Padgett, Science (New York, N.Y.) **340**, 844 (2013).
[17] P. Ryczkowski, M. Barbier, A. T. Friberg, J. M. Dudley, and G. Genty, Nature Photonics (2016), 10.1038/nphoton.2015.274.
[18] G. Brida, M. Genovese, and I. Ruo Berchera, Nature Photon. **4**, 227 (2010), arXiv:1004.1274.
[19] P. A. Morris, R. S. Aspden, J. E. C. Bell, R. W. Boyd, and M. J. Padgett, Nature communications **6**, 5913 (2015), arXiv:1408.6381.

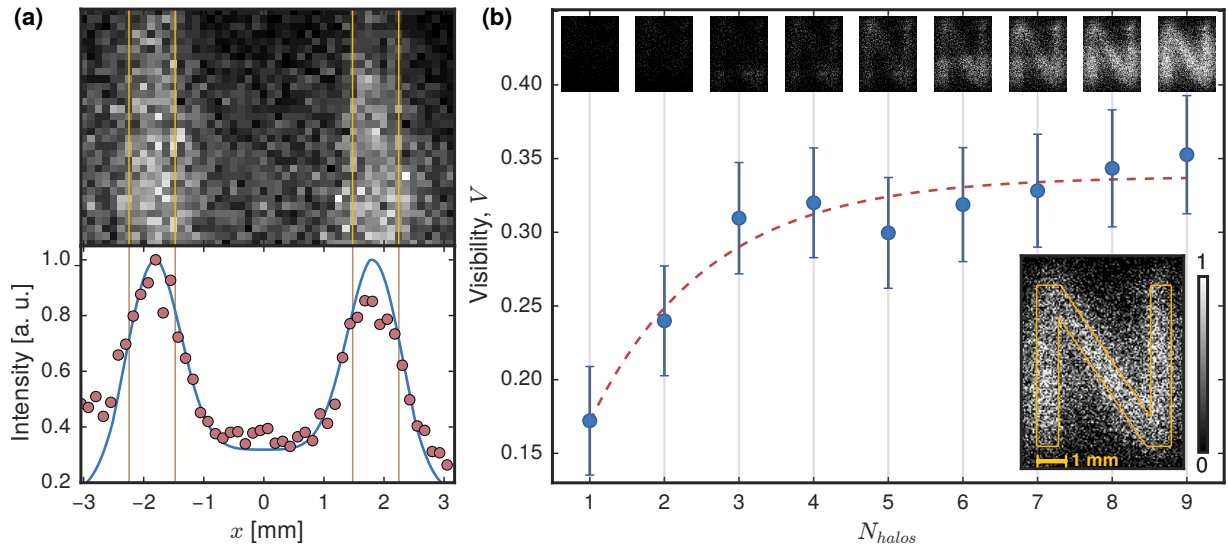

FIG. 4. **Ghost image resolution and visibility**. (a) Ghost image of the vertical bars of the "U" (yellow bars). The image (top) is integrated vertically (bottom, circles) to yield an intensity, which is fitted with the convolution with the PSF [29] (solid line). This yields a width (representing the imaging resolution) of 0.40 mm. (b) The visibility ($V = (I - B)/(I + B)$) is shown as a function of the number of halos accumulated to form the image ($N_{halos}$). $I$ is the mean intensity within the shape to be imaged (yellow "N", as shown in the inset), while $B$ is the mean intensity outside the "N". Each plotted image is the result of accumulating reconstructed images from different $s$-wave halos. The dashed curve is a guide to the eye, while error bars show the standard error of the mean across the image pixels.

[20] R. S. Bennink, S. J. Bentley, and R. W. Boyd, Physical review letters **89**, 113601 (2002).
[21] R. S. Bennink, S. J. Bentley, R. W. Boyd, and J. C. Howell, Physical review letters **92**, 033601 (2004).
[22] F. Ferri, D. Magatti, a. Gatti, M. Bache, E. Brambilla, and L. a. Lugiato, Physical Review Letters **94**, 2 (2005), arXiv:0408021 [quant-ph].
[23] T. Jeltes, J. M. McNamara, W. Hogervorst, W. Vassen, V. Krachmalnicoff, M. Schellekens, a. Perrin, H. Chang, D. Boiron, a. Aspect, and C. I. Westbrook, Nature **445**, 402 (2007), arXiv:0612278 [cond-mat].
[24] S. S. Hodgman, R. G. Dall, A. G. Manning, K. G. H. Baldwin, and A. G. Truscott, Science **331**, 1046 (2011), http://science.sciencemag.org/content/331/6020/1046.full.pdf.
[25] Dall R. G., Manning A. G., Hodgman S. S., RuGway Wu, Kheruntsyan K. V., and Truscott A. G., Nat Phys **9**, 341 (2013).
[26] P. L. Kapitza and P. A. M. Dirac, Mathematical Proceedings of the Cambridge Philosophical Society **29**, 297 (1933).
[27] Y. B. Ovchinnikov, J. H. Müller, M. R. Doery, E. J. D. Vredenbregt, K. Helmerson, S. L. Rolston, and W. D. Phillips, Phys. Rev. Lett. **83**, 284 (1999).
[28] R. Dall and A. Truscott, Optics Communications **270**, 255 (2007).
[29] "Supplementary Materials,".
[30] A. G. Manning, R. I. Khakimov, R. G. Dall, and A. G. Truscott, Nature Physics **11**, 539 (2015).

## SUPPLEMENTARY MATERIALS

### Experimental Apparatus

The He* BEC is initially trapped in a bi-planar quadrupole Ioffe configuration magnetic trap [28], with harmonic frequencies of $\{\omega_x, \omega_y, \omega_z\}/2\pi \approx \{15, 25, 25\}$ Hz and a bias field of $B_0 = 1.31(\pm 0.01)$ G along $x$-axis. The 80 mm diameter multichannel plate and delay line detector is located $\sim$850 mm below the trap center (416 ms fall time), with a spatial resolution in $x$ and $y$ of $\sim$120 $\mu$m, a temporal resolution along $z$ of $\sim$2 ns and a quantum efficiency of $\sim$10%.

### Bragg diffraction

Similarly to our previous work [30] we employ the same laser beams for both Raman and Bragg pulses, only changing the relative frequency detuning of the waveforms, which is set by the bias $B_0$ and geometrical angle between the beams (90°). The laser is blue detuned by 2 GHz from the $2^3P_0$ state. The Raman transition in the beginning of the sequence also results in the cloud acquiring a momentum change of $\sqrt{2}\hbar k_0 \mathbf{e_z}$, which sets the center-of-mass of the cloud in downward motion. The duration of the outcoupling Raman pulse is $t_\pi = 1.8$ $\mu$s, with $\sim$95% transfer efficiency. It is then immediately followed by a 1.4 $\mu$s Bragg pulse. The maximum intensity of each laser beam is $\sim$450 mW/mm$^2$ for the Raman pulse and $\sim$30 mW/mm$^2$ for the Bragg pulse. Each of the pulse sequences is modulated with an overall Gaussian envelope to control Fourier-broadening. Thus, the broadening of the Raman pulse was optimised to maximise the transfer to the magnetically insensitive sublevel $m_J = 0$, while the Bragg pulse was set to choose a momentum transfer to be in the Kapitza-Dirac regime and therefore to maximise the number of Bragg orders generated ($L$).

In our experiment we have achieved $L = 12$ orders, consequently producing $L - 1 = 11$ $s$-wave halos between the adjacent orders, as pictured in Fig. 2b. In principle, by increasing the laser power of the Bragg beams, one could raise $L$ even further, generating more halos with lower density and therefore possessing higher correlations. It is, however, optimal to have $L = 12$ for the ghost imaging configuration in our experiment because of the relatively high absolute velocity of the $^4$He* atoms in the higher Bragg orders, i.e. for $l = \pm 6$ it is $v_z = \pm 78$ cm/s. There are two problems which arise for $|l| > 6$ in our setup: (i) fast travelling downwards kicked halos do not expand enough to cover the object mask before falling onto the detector (Fig. 2c), and (ii) upwards kicked halos will be lost through hitting the top of the vacuum chamber.

It is also interesting to note that we observe larger-diameter halos from the non-adjacent orders $(l, l+2)$, which are reconstructed only partially since they expand beyond the detector size. In principle, $g^{(2)}$ peaks should be higher for these larger halos, since the scattering modes overlap less [4]. However, the population of these halos is too low to make them usable here.

Note that the non-zero halo thickness $\delta k$ of the halo is due to uncertainty limited broadening from the finite size of the condensates, as well as mean-field interactions in the BEC during expansion of the halo. In our experiments we observe $\delta k/k_r = 0.034(\pm 0.003)$.

### Correlation function

For independent $s$-wave halos originating from different Bragg orders (Fig. 2b) we construct a second-order normalised cross-correlation function:

$$g^{(2)}(\Delta \mathbf{k}) = \frac{\int d^3\mathbf{k}\, G^{(2)}(\mathbf{k}, -\mathbf{k} + \Delta \mathbf{k})}{\int d^3\mathbf{k}\, \langle n(\mathbf{k})\rangle \langle n(-\mathbf{k} + \Delta \mathbf{k})\rangle}. \quad (1)$$

We find that the rms widths of the $g^{(2)}$ are approximately constant across all Bragg orders. Note that the correlation lengths we measure are actually a convolution of real correlation length with the detector resolution, which is $\sim 120\mu m$.

The error estimate in the numerical evaluation of $g^{(2)}$ can be expressed as a standard deviation of the atom count frequencies (which represent the numerator of Eq. 1) normalised to the same denominator. The error bars on Fig. 3 represent a $2\sigma$ confidence interval.

Because of the spherical symmetry of the $s$-wave halo in momentum space, it is convenient to operate with atomic momenta, while expressing the final results in spatial units. Knowing the time-of-flight to the detector $T_f$, we convert atom velocities $\Delta \mathbf{v} = \hbar \Delta \mathbf{k}/m$ to spatial coordinates in the detector plane: $\mathbf{r} = T_f \Delta \mathbf{v}$, where $m$ is the mass of a $^4$He* atom. This conversion yields the spatial co-ordinate representation for $g^{(2)}$ in Fig. 3.

### Ghost image

The target to be imaged was made out of laser cut 0.4 mm thick stainless steel sheet. A microscope image of the object used for the ghost imaging is shown in Extended Data Fig. 5.

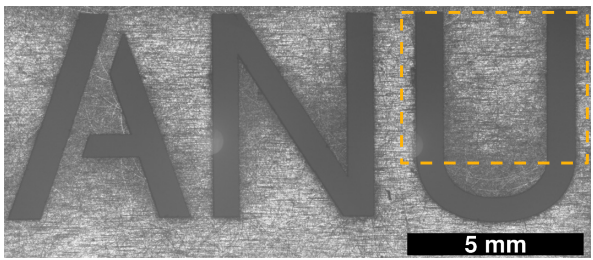

FIG. 5. **The object**. Microscope image of the mask used to create the ghost image. The region indicated by the dashed line forms the vertical bars shown in Fig. 4a of the main text, which was used to determine the imaging resolution.

The ghost image is reconstructed in $k$-space using coincidence-counting between atoms in the multi-pixel port $n_M$ and the bucket port $n_B$:

$$\widetilde{C}(k_x, k_y) = n_M(k_x, k_y, k_z) \otimes n_B(k_z). \quad (2)$$

Next, analogously to $g^{(2)}$, we convert the image $\widetilde{C}(k_x, k_y)$ to spatial coordinates $C(x, y)$ in the detector plane $z = 0$. This image is shown in Fig.2c. The correlator in Eq. 2 accepts two counts from $n_M$ and $n_B$ as a valid coincidence if they occur within the correlation length $\sigma_z$ in the $z$ (time) direction, which is given by the $g^{(2)}$ width of 0.37mm (see Fig. 3).

### Image resolution and visibility

We estimate the resolving power of our ghost imaging configuration by taking the ghost image of the vertical bars from the "U" part of the "ANU" mask (Fig. 5), and fitting it with the convolution $Q(x) = b * h$ of the real object shape $b(x)$ and the Gaussian point spread function (PSF) $h(x) = B_0 + C \exp(-x^2/2s_x^2)$. Therefore, using the rms width $s_x$ of the PSF as one of three free parameters of the fit, we can compare the resolution of the ghost image along the $x$-direction with the correlation length along that axis. As stated in the main text, we find $s_x = 0.40(\pm 0.03)$ mm, which is in good agreement with the rms width of $\sigma_x = 0.43(\pm 0.01)$ mm extracted from $g^{(2)}$.

We characterise the visibility of the ghost image by comparing the average image intensity $I$ to the average background $B$:

$$V = \frac{I - B}{I + B}, \quad (3)$$

where the image intensity $I$ is calculated within the region which matches the object mask. This region is determined by manually overlapping an "N" shaped region with the exact dimensions of the mask on top of the ghost image. Extended Data Fig. 6 shows the ghost image visibility for images from each individual halo as a function of the average atom number $\langle N_a \rangle$ in that halo. The inset shows the corresponding object (the letter "N") and its ghost image for the halo with the highest number of counts. Because of the different center-of-mass velocities, the halos with the smallest time-of-flight only partially cover the mask. The first two halos to arrive onto the detector $(+6, +5)$ and $(+5, +4)$ have a small halo population $\langle N_a \rangle$ as well as a minimal overlap with the mask, which results in a weak ghost image signal relative to the background and hence $V < 0$. Therefore, these halos were not used to produce the cumulative ghost image shown in Fig. 4 of the main text. It is important to note that Eq. (3) gives high visibility values (many counts within the object region and a small background), – even if the halos do not cover the object completely, and thus do not generate a full representation of the entire image. However, we have used this visibility definition to be consistent with the definition employed in the photon ghost-imaging literature [19].

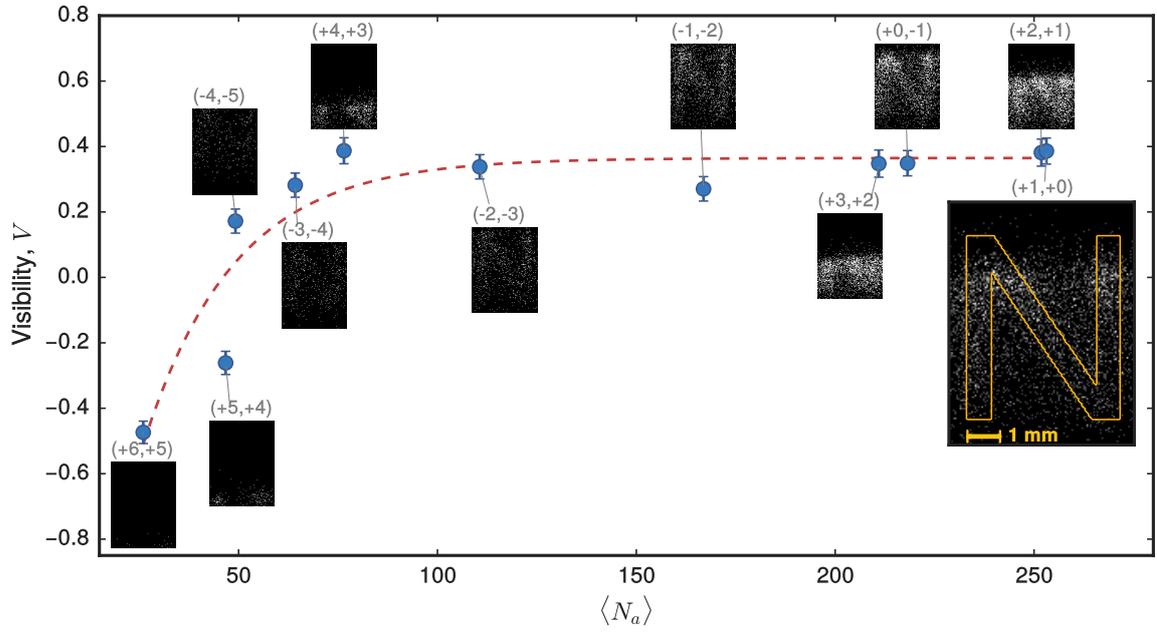

FIG. 6. **Ghost image visibility**. Visibilities (dots) for images (insets) reconstructed from each individual halo with different average number of atoms $\langle N_a \rangle$. Bragg orders producing the halos are labelled as $(l+1, l)$. The dashed curve is a guide to the eye. Error bars represent the standard error of the mean associated with the variances of the pixel values contributing to $I$ and $B$.